\documentclass[pra,amssymb,showpacs]{revtex4}

\usepackage{graphicx}
\usepackage{amssymb,amsmath,amsbsy} 
\usepackage{appendix}

\def\vp{\varphi}
\def\th{\theta}

\def\d{\partial}

\def\a{\alpha}

\def\ve#1{{\bf #1}}

\def\En{\hat{\mathcal E}}
\def\be{\beta_\ell}

\def\EE{\hat{\ve E}}
\def\HH{\hat{\ve H}}

\def\b0{\hat b_{\ell0}}

\def\bu0{\hat{ \underline{ b_0}}}

\def\ch{{\rm cosh}}
\def\sh{{\rm sinh}}
\def\th{{\rm tanh}}
\def\ath{{\rm arctanh}}

\def\Y{\tilde Y}

\begin{document}

\title{On the quantum Coulomb field}

\author{Bogdan Damski$^1$ and Piotr Marecki$^2$} 
\affiliation{$^1$Los Alamos National Laboratory, Theoretical Division, MS B213, Los Alamos, NM, 87545, USA \\
$^2$Fakult\"at f\"ur Physik, Universit\"at Duisburg-Essen, Lotharstrasse 1, 47057 Duisburg, Germany} 
\begin{abstract}
The quantum theory of the Coulomb field has been developed  by Staruszkiewicz in the long series of papers.  
This theory  explains  the universality and quantization of the electric charge observed in Nature. 
Moreover, the efforts have been made to determine the value of the elementary charge
from its  mathematical structure.
Nonetheless, no other immediate applications of this theory have been proposed. We make such an attempt by 
(i) considering the classical energy operator and defining its counterpart in the quantum theory of the Coulomb field;
(ii) determining the eigenstates of the energy operator and assigning energy to the excitations of the theory; and 
(iii) proposing a simple theoretical scheme to estimate   the effect of the quantum fluctuations 
of the Coulomb field on the energy levels of hydrogen-like atoms. We argue that the recent experimental advances 
in hydrogen and muonic-hydrogen spectroscopy may
provide the unique window of opportunity for the verification 
of the Staruszkiewicz's theory.
\end{abstract}
\maketitle

\section{Introduction}
 
The theory of the quantum Coulomb field (QCF) has been proposed by Staruszkiewicz 
\cite{Annals1989,Acta1992,Acta1995} 
(see also Ref. \cite{AH2005} for the discussion of this theory from  a different
angle and Refs. \cite{GZ1980,Morchio1983471,AH2008} for other approaches to the quantization of
long-range fields). It supplements standard
quantum electrodynamics (QED)  by describing {\it long-range}
quantum fluctuations of the Coulomb field \cite{Acta1998}.
The principal need for the development of this theory 
is that even though the electric charge of every particle type \emph{could} be arbitrary
in the QED framework, all electric charges are multiples of a single quantum with a stunning accuracy \cite{Acta2002}. 

The studies of the quantum Coulomb field have been
centered on charge quantization and universality as well as on the
extensive
search for a distinguished value of the fine structure constant (see e.g. Ref. \cite{Foundations2002}). 
While the former 
efforts have succeeded, the Staruszkiewicz's theory predicts that charged
particles carry an integer multiple of the same elementary charge, the latter
efforts are still ongoing. Indeed, even though it was found that 
the mathematically distinguished interval of the allowed values of $\a$ is \mbox{$0<\alpha<\pi$}, the hunt for the
theoretical determination of the exact value of the fine structure constant \mbox{$\alpha=1/137.036\dots$}
is unfinished. We remark that  to the best of our knowledge 
the Staruszkiewicz's theory provides the first theoretical framework in which such a  hunt is at all possible.

Our goal is to extend the QCF theory to study   
its experimentally-relevant   implications. 
We discuss the electric and magnetic field  operators of the
QCF, define a set of distinguished  states of the QCF  that might be carried by 
charged particles, and develop a simple theory 
providing a basis for the  quantitative studies of the shifts of the 
energy levels of atoms resulting from the quantum fluctuations of the Coulomb field.
The possibility that there may be additional level shifts absent in standard
QED is strongly suggested by the puzzling
recent precision spectroscopy experiment on muonic hydrogen \cite{PohlNature2010}.

\section{Basics of the quantum Coulomb field theory}
In this section,  we will briefly summarize the QCF theory of Staruszkiewicz. 
The central object of the theory is the phase field $S(x)$. 
It may be derived on the classical level in two ways. 

Such a degree of freedom is always present for charged fields.
This is seen by writing the action for the electromagnetic field and its
source in 
the form
\begin{equation}
-\frac{1}{16\pi}\int d^4x\, F^{\mu\nu}F_{\mu\nu} + \int d^4x\, \mathcal L[e A_\mu+\d_\mu S,\ldots], 
\label{Actione}
\end{equation}
where the phase $S$ of the charged matter field
appears in the gauge-invariant combination $e A_\mu+\d_\mu S$, 
while the dots stand for other degrees of freedom of the charged matter field, e.g.,
its real amplitude, 
relative phases between the different components of a spinor, etc.

To illustrate this point, one can consider the simplest example of a field
theoretical system with the electric charge: the Klein-Gordon theory
discussed in this context in Ref. \cite{Acta1990}. Its gauge-invariant action reads 
$\protect{-\frac{1}{16\pi} \int d^4x F^{\mu\nu}F_{\mu\nu} +
\frac{1}{2} \int d^4x |(\partial^\mu+ieA^\mu)\phi|^2-m^2 |\phi|^2}$.
Writing the Klein-Gordon field $\phi$ as $R\exp(iS)$ one finds that this
action equals  
$$
-\frac{1}{16\pi} \int d^4x F^{\mu\nu}F_{\mu\nu} +\frac{1}{2}\int
d^4x \left[  \partial^\mu R\partial_\mu R + R^2 (\partial_\mu S + eA_\mu) (\partial^\mu S + eA^\mu)
-m^2R^2\right],
$$
in agreement with Eq. (\ref{Actione}). In this particular case, the dots in the
argument of  $\mathcal L$ stand for $R$ (the real  amplitude of the charged matter field) and its
derivatives.

Coming back to the discussion of Eq. (\ref{Actione}), we note that its variation 
with respect to $A_\mu$  defines the current $j_\mu$ to be equal to 
$\frac{1}{4\pi}\partial^\nu F_{\mu\nu}$.
The variation of the action with respect to $S$ leads to the charge conservation law: $\partial_\mu j^\mu = 0$.
The momentum canonically conjugated with the phase field $S$ is
\begin{equation*}
\pi_S = \frac{\partial \mathcal L}{\partial\left(\frac{\partial S}{\partial t}\right)} =
- \frac{j_0}{e}.
\end{equation*}
The canonical quantization $[\hat S(x),\hat \pi_S(y)]_{x^0=y^0}=i\delta({\bf x}-{\bf y})$ results in 
\begin{equation*}
[\hat{j_0}(x),\hat S(y)]_{x^0=y^0} = ie \delta({\bf x}-{\bf y}).
\end{equation*}
Integrating it over the hyperplane $x^0=y^0$, one obtains
\begin{equation}
 [\hat Q,\hat S(y^0,{\bf y})]=ie,
\label{Weinberg}
\end{equation}
where $\hat Q=\int d^3x\,\hat{j_0}$ is the  charge operator, 
$e=1/\sqrt{137.036\dots}$ stands for the unit of the electric charge, 
and $\hbar=c=1$. This interesting relation, however, is rather 
useless unless further input is provided about the phase field $\hat S(x)$. This is 
done in the following way.

The electric charge can be also  defined by looking only at the electromagnetic field. 
 In order to find the minimal framework for such a consideration, one 
investigates the following scalar functional of the electromagnetic potential $A_\mu(x)$: 
\begin{equation*}
\mathfrak s(x)=-ex^\mu A_\mu(x).
\end{equation*}
It can be shown that it characterizes the electromagnetic fields of the Coulomb type completely.
Namely, all electric and magnetic fields falling off as $1/r^2$ can 
be uniquely expressed as 
the appropriate derivatives of $\mathfrak s(x)$ through 
the relation $-ex^\nu F_{\mu\nu} = \partial_\mu \mathfrak s(x)$ (see e.g. Ref. \cite{Acta2002}).  
Moreover, $\mathfrak s(x)$ satisfies the d'Alembert equation: $\Box\mathfrak s = 0$.
 
In Ref. \cite{Annals1989}, Staruszkiewicz formulated 
a complete field-theoretical system by reflecting on the Gauss law. He 
identified $S(x)$ 
(the property of the charged matter) with $\mathfrak s(x)$ (the property of the electromagnetic field), 
and used the commutation relation (\ref{Weinberg}) 
as the basis of the quantum field theory of the Coulomb field. 
Note that the identification of a degree of freedom of matter with some degree of freedom of 
the electromagnetic field is necessary if matter is required to carry a true electric 
charge even in asymptotic future/past (see e.g. Ref. \cite{AH2008}).

The quantization of the phase field proceeds in two steps. 
Firstly, as the consequence of both the d'Alembert equation satisfied by $S(x)={\mathfrak s}(x)$ and
the required fall-off condition for $S(x)$, one obtains the appropriate field operator \cite{Annals1989}:
\begin{equation*}
\hat S(x) =\hat S_0 - e \hat Q\, \th(\psi) +
\sum_{\ell=1}^\infty\sum_{m=-\ell}^\ell\left[\hat c_{\ell m}
f_{\ell m} + {\rm h.c.}\right],
\end{equation*}
$$
 f_{\ell m}(\psi,\theta,\varphi) =
 \left[
{_2F_1}\left(-\tfrac{\ell+1}{2},\tfrac{\ell}{2}\, ;\tfrac{1}{2}\, ;{\rm \th}^2\psi\right)\frac{G_\ell}{2}
- \frac{i}{G_\ell}\, \th(\psi) 
  {_2F_1}\left(-\tfrac{\ell}{2} ,\tfrac{\ell+1}{2};\tfrac{3}{2};{\rm \th}^2\psi\right)\right]
 Y_{\ell m}(\theta,\varphi),\nonumber
$$
where $(r,\theta, \varphi)$ are the spherical coordinates 
around the charge, $\psi=\ath(t/r)$ and 
$$
G_\ell =\sqrt{\frac{\ell}{\ell+1}}\frac{\Gamma\left(\tfrac{\ell}{2}\right)}{\Gamma\left(\tfrac{\ell+1}{2}\right)}.
$$ 
The phase field $\hat S$ is  quantized {\it outside} of
the light cone $x_\mu x^\mu=t^2-{\bf x}^2=0$, i.e., for $r>|t|$.
The non-vanishing commutators are
$$
[\hat c_{\ell m},\hat c^\dag_{\ell'm'}]=4\pi e^2\delta_{\ell\ell'}\delta_{mm'}, \  \ [\hat Q,\hat S_0]=ie.
$$

As the second step of quantization, the representation space for the theory is defined to 
be the Fock space constructed upon the ``vacuum'' state $|0\rangle$, 
which is supposed to be annihilated by all $\hat c_{\ell m}$  and $\hat Q$ 
(the last relation being highly non-trivial; see also \cite{footnote1}). 

The ``simplest'' state carrying $n$ units of the electric charge, 
$|n\rangle$, is created from vacuum via $\exp(-in \hat S_0)$:
\begin{equation}
|n\rangle=\exp(-in \hat S_0)|0\rangle, \ \ \hat Q|n\rangle= ne |n\rangle, \
\langle n|m\rangle = \delta_{nm}.
\label{vacuum}
\end{equation}
We will refer to $|n\rangle$ as the $n$-charged vacuum state as it is  
annihilated by all $\hat c_{\ell m}$. One should remember that $|n\rangle$ is
a short cut for $|n;\{n_{\ell m}\}\rangle$, where $n_{\ell m}$, 
the occupation of the $(\ell,m)$ modes, is set to zero. The $(\ell,m)$ modes 
are responsible for the  global angular distortions of the Coulomb field.
Another distinguished
charged state of the QCF was found in Ref. \cite{AstarBound}.

Finally, we would  like to draw  attention of the reader 
to the fact that independently of the beautiful physical relevance 
attached to the field $\hat S(x)$, the quantum theory of this field is a perfectly well-defined, highly interesting, and well worked-out quantum field theory in $2+1$ dimensional de Sitter spacetime (see e.g. the recent non-trivial
results on the structure of the boost operators and the spectral decomposition of the vacuum state \cite{Acta2004,Reports2009}).

\section{Fluctuations of the quantum Coulomb field}

It is convenient for our calculations to switch to the following notation.
For $m=0$, we define
\begin{equation*}
\hat c_{\ell0} = \sqrt{4\pi}ie\, \hat b_{\ell0}, \ \ \Y_{\ell0}(\theta,\varphi) = Y_{\ell0}(\theta,\varphi), 
\end{equation*}
while for $m>0$, we define 
$$
\hat c_{\ell,\pm m} = \sqrt{2\pi}ie\left(\hat b_{\ell m} \pm \hat b_{\ell,-m}\right), \  \
\Y_{\ell,\pm m}(\theta,\varphi) = \frac{Y_{\ell m}(\theta,\varphi)\pm Y_{\ell,-m}(\theta,\varphi)}{\sqrt{2}}.
$$
The redefined operators satisfy
$[\hat b_{\ell m},\hat b^\dag_{\ell'm'}]=\delta_{\ell\ell'}\delta_{mm'}$ and $[\hat b_{\ell m},\hat b_{\ell'm'}]=0$.

From a given $\hat S(x)=\hat{\mathfrak s}(x)$ the corresponding electromagnetic Coulomb fields can be computed using
\begin{equation*}
\hat A^\mu(x) = -\frac{1}{e}\frac{x^\mu}{xx} \hat S(x).
\end{equation*}

In the standard spherical orthonormal tetrad $({\bf N},{\bf \Theta}, {\bf \Phi})$:
\begin{align*}
{\bf N}&=(\sin\theta\cos\varphi,\sin\theta\sin\varphi,\cos\theta),\\
{\bf \Theta}&=(\cos\theta\cos\varphi,\cos\theta\sin\varphi,-\sin\theta),\\
{\bf \Phi}&= (-\sin\varphi,\cos\varphi,0),
\end{align*}
we find
\begin{align*}
\hat {\bf E} &= -\frac{1}{er}\left[{\bf N}\frac{\partial}{\partial t}
+\frac{t}{r^2-t^2}
\left(\frac{\bf \Phi}{\sin\theta}\frac{\partial}{\partial\varphi} + 
      {\bf \Theta}\frac{\partial}{\partial\theta}\right)
\right]\hat S,\\
\hat {\bf H} &= \frac{1}{e}\frac{1}{r^2-t^2}
\left(\frac{\bf \Theta}{\sin\theta} \frac{\partial}{\partial\varphi} -{\bf \Phi}\frac{\partial}{\partial\theta}\right)\hat S.
\end{align*}

Within the theory of Staruszkiewicz, the Coulomb electromagnetic fields 
are given only in the spacetime region $|t|<r$  (the theory is originally
formulated in spatial infinity outside of the  light cone,  where
this condition does not cause problems; we extrapolate the fluctuations from 
the spatial infinity to near the charge center assuming homogeneity of the
Coulomb field).
This reflects the lack of a complete theory of charged matter accompanied by such fields 
(for attempts in this direction see Ref. \cite{AH2008}). 
In this paper, we circumvent that difficulty by dealing exclusively with effects on the 
surface $t=0$, 
for which the theory gives complete information about the electromagnetic fields 
(apart from the singular point $r=0$): 
\begin{align}
\label{Efluc}
\EE(0,{\bf x}) &= \left[\hat
Q-\sqrt{4\pi}\sum_{\ell=1}^\infty\sum_{m=-\ell}^\ell
\left(\frac{\Y_{\ell m}}{G_\ell}\hat b_{\ell m}+{\rm h.c.}\right)\right]\frac{\bf N}{r^2}, \\
\HH(0,{\bf x}) &= \frac{\sqrt{\pi}}{r^2}
\left[\frac{{\bf\Theta}}{\sin\theta}\frac{\partial}{\partial\varphi}-{\bf\Phi}\frac{\partial}{\partial\theta}\right]
\sum_{\ell=1}^\infty \sum_{m=-\ell}^\ell
(i\Y_{\ell m}G_\ell \hat b_{\ell m}+{\rm h.c.}), 
\label{Hfluc}
\end{align}
Several remarks are in order now. 

First,  these expressions show
 that there is a fluctuating electric and magnetic field around a
charge: an expected feature of the quantum Coulomb field. 
The amount of fluctuation depends on the quantum
state of its Coulomb field. Physically sound  states describing the QCF of 
a charge at rest  
are those reproducing the classical result:
\begin{equation}
\langle \,\hat {\bf E}(0,{\bf x})\,\rangle=ne\frac{\bf N}{r^2}, \ \ \langle\,\hat {\bf H}(0,{\bf x})\,\rangle=0.
\label{class}
\end{equation}
This condition, however, is satisfied by the infinite number of states. For
example, by any single Fock state $|n, \{n_{\ell m}\}\rangle$, where 
$n_{\ell m}$ is an arbitrary non-negative integer. It is therefore 
fundamentally important to ask which quantum state of the Coulomb field is carried 
by a charged particle, say an electron or a proton ? 

Second, the amount of the fluctuation of the quantum Coulomb field should be
experimentally measurable through the  studies of the level shifts of atoms (especially
hydrogen-like for which the techniques of precision spectroscopy work best
\cite{SpectroscopyAll,HaenschRMP2006}; we will come back to this point later). 
This is a simple conclusion well founded on the
success of QED to explain the Lamb shift. The Lamb shift, originally the splitting of the 
$2S_{1/2}$ and $2P_{1/2}$ levels in hydrogen, results from the coupling between
the orbiting charge and the fluctuating vacuum field
\cite{BethePhysRev,Bethe,MilonniBook1993}. 

Third, one easily sees from (\ref{Efluc})  that 
the relative fluctuation of the QCF with respect
to its classical counterpart is (i) independent of the distance from the charge center as  
both decay as $1/r^2$ and (ii) decays as the inverse of the charge carried by 
the QCF (fluctuations are charge independent while the classical field goes linearly with charge). 
The (i) effect is in stark contrast 
to  what one finds in QED, where the relative modification of the classical Coulomb field 
by virtual electron-positron pairs is  short-ranged \cite{MilonniBook1993,EidesBook2007}. 

\section{Energy associated with the quantum Coulomb field}

Having realized that there are many quantum states of the Coulomb field that
are
physically allowed, we propose a simple physically-motivated approach 
allowing for finding a set of distinguished states of the QCF. Namely, we
propose to ``label'' different configurations of the QCF by an energy 
associated with them. To this aim, we define electromagnetic energy operator  
on the surface $t=0$:
\begin{equation}\label{en_op}
\En_C=\frac{1}{8\pi}\int d^3x\,:\EE(0,\ve x) \EE(0,\ve x)+\HH(0,\ve x)\HH(0,\ve x):,
\end{equation}
where $::$ stands for normal ordering \cite{footnote2}. 

In classical electrodynamics the expression \eqref{en_op} corresponds to the energy of field configurations. 
In the QCF theory states reproducing on average the classical Coulomb field do exist, 
and we feel it justified to claim that their energy should be computed from the quantum
version of the classical energy functional. 
However, since $\EE(0,{\bf x})$ and $\HH(0,{\bf x})$ scale  as $1/r^2$, 
this operator has problems 
due to the divergence of the integral at $r=0$. Moreover, it 
does not have the interpretation of the generator of the time-translation symmetry (the Hamiltonian). 
We will discuss the former problem below and mention here  that it is our assumption to 
assign the energy to the excitations of the QCF through the operator (\ref{en_op}). 

Putting operators (\ref{Efluc}) and
(\ref{Hfluc}) into the above expression, we obtain 
\begin{align}
\En_C &= \frac{\varepsilon_c}{e^2}\hat Q^2 + 
\frac{\varepsilon_c}{e^2} \sum_{\ell=1}^\infty \gamma_\ell \hat{\cal E}_\ell,
 \label{A} \\
\hat{\cal E}_\ell &=
\b0^\dag\b0+\frac{\be(-1)^\ell}{2}(\b0^2+\b0^{\dag2}) +
\sum_{m=1}^\ell  \left[
\hat b_{\ell m}^\dag \hat b_{\ell m}+  
 \hat b_{\ell,-m}^\dag \hat b_{\ell,-m} 
+  (-)^{m+\ell}
\frac{\beta_\ell}{2} (\hat b_{\ell m}^2-\hat b_{\ell,-m}^2+{\rm h.c.})\right].
 \label{EEE}
\end{align}
We introduced above
$$
\beta_\ell= \frac{4-G_\ell^4\ell(\ell+1)}{4+G_\ell^4\ell(\ell+1)},\ \ 
\gamma_\ell=\frac{2}{G_\ell^2}+\frac{G_\ell^2\ell(\ell+1)}{2},
$$
and  
\begin{equation}
 \varepsilon_c=\frac{e^2}{8\pi}\,\int \frac{d^3 x}{r^4}.
\label{eps_c}
\end{equation}
The latter corresponds to the classical electromagnetic energy 
of a unit charge at rest. We do not propose any new solution to 
the obvious lack of convergence of this integral at $r=0$. On general grounds, however, 
we expect that in any such solution the parameter $\varepsilon_c$ should be of the order 
of the electron's rest mass $m_e$. This is 
in line with the century long efforts to 
associate the  mass of an electron with the energy of its field  \cite{elmass} 
(see also Ref. \cite{DurrScience2008} for the relevant hadronic example). 
We shall also argue below that the actual spectroscopic consequences of the quantum theory of Coulomb 
field will be in a good position to determine the numerical value of $\varepsilon_c$.

To look for the  spectrum of the electromagnetic energy operator, we diagonalize 
\eqref{A}.
This is done by applying the Bogolubov transformation
\begin{equation}
\hat B_{\ell m}={\rm \ch}(u_{\ell m})\hat b_{\ell m}-{\rm \sh}(u_{\ell m})\hat b_{\ell m}^\dag, 
\label{B}
\end{equation}
where  assuming that $m>0$ 
\begin{equation}
u_{\ell0}=\frac{(-)^{\ell+1}\text{\ath}(\be)}{2},  
u_{\ell, \pm m} = \frac{\mp(-)^{m+\ell}\text{\ath}(\be)}{2}.
\label{u}
\end{equation}
The transformation (\ref{B}) can be easily expressed as the explicit function
of $\ell$ employing
$$
{\rm \ath}(\beta_\ell) = \ln\frac{2}{G_\ell^2\sqrt{\ell(\ell+1)}}.
$$
Finally, we note that $[\hat B_{\ell m},\hat B_{\ell' m'}^\dag]=\delta_{\ell\ell'}\delta_{mm'}$
and $[\hat B_{\ell m},\hat B_{\ell' m'}]=0$ hold for the redefined operators.

Putting Eqs. (\ref{B}) and  (\ref{u}) into Eqs. (\ref{A}) and (\ref{EEE}), we find 
\begin{equation}
\En_C =\hat Q^2 \frac{\varepsilon_c}{e^2} +
\frac{\varepsilon_c}{2e^2} \sum_{\ell=1}^\infty \gamma_\ell\left(\sqrt{1-\beta_\ell^2}-1\right)(2\ell+1)
+\frac{2\varepsilon_c}{e^2}\sum_{\ell=1}^\infty
\sum_{m=-\ell}^\ell\sqrt{\ell(\ell+1)} \hat B_{\ell m}^\dag\hat B_{\ell m},
\label{Diag}
\end{equation}
where we have employed the  identity 
$\gamma_\ell\sqrt{1-\beta_\ell^2}=2\sqrt{\ell(\ell+1)}$ to simplify the result. 

The first term in (\ref{Diag}) recovers the classical electromagnetic energy 
of a charge at rest. The second one is the contribution of the zero point 
modes (see also \cite{footnote2})
\begin{equation}\label{shift}
\frac{\varepsilon_c}{2e^2} \sum_{\ell=1}^\infty
\gamma_\ell\left(\sqrt{1-\beta_\ell^2}-1\right)(2\ell+1)\approx -5\times10^{-2} \frac{\varepsilon_c}{e^2}.
\end{equation}
The third term in Eq. (\ref{Diag}) provides the spectrum of the electromagnetic 
energy operator, which is degenerate for all $m$ in every $\ell$ sector.
We define the ground state of $\En_C$,
say $|G\rangle$, as the state that is annihilated by all $\hat B_{\ell m}$ operators.
For generality, we assume that it carries $n$ quanta of the electric
charge: $\hat Q|G\rangle = ne|G\rangle$.
The excited states are created in the $n$-charged sector by acting (an arbitrary number of times)
$\hat B_{\ell m}^\dag$ on  $|G\rangle$. It is worth to mention that the
excitation gap  to the $\ell$ sector equals
\begin{equation}
\frac{2\varepsilon_c\sqrt{\ell(\ell+1)}}{e^2},
\label{gap}
\end{equation}
which is a huge number assuming that indeed $\varepsilon_c={\cal O}(m_e)$.
Given the physical interpretation of the electromagnetic energy operator and
an expected huge gap in the excitation spectrum,
we propose  the ground state $|G\rangle$ as the natural candidate 
for representing the quantum state of the Coulomb field of a stable charged
particle (e.g. an electron or a proton \cite{footnote3}). 
Note that both the charged vacuum state $|n\rangle$  and the 
ground state $|G\rangle$ are spherically symmetric. The 
expectation value of the energy operator is lowest in the state $|G\rangle$ (the state 
$|n\rangle$ is not an eigenstate of this operator).
In what follows, we  describe some of its basic properties.

First, we note that the expectation value of the electric and magnetic field
operators, calculated in the ground state $|G\rangle$, reproduces the classical result (\ref{class}).
States supposed to correspond to moving particles can be constructed because the generators of the 
boosts  are explicitly known (see Refs. \cite{Wrobel,Rostworowski} and the Appendix). 
For example, the  quantum state of the Coulomb field of the charge moving 
in the $+z$ direction with velocity $v=$\th$\lambda$  is 
$$
|G,\lambda\rangle=\exp(-i\lambda\hat M^{03})|G\rangle,
$$
where $\hat M^{03}$ is the boost operator (\ref{AM03}).

We expect that the averages, 
\begin{equation*}
\langle G,\lambda|\hat {\bf E}(0,{\bf x})|G,\lambda\rangle, \ \ 
\langle G,\lambda|\hat {\bf H}(0,{\bf x})|G,\lambda\rangle,
\end{equation*}
computed as the
functions of {\bf x} on the surface $t=0$, are equal to the classical fields
of a moving charge crossing this surface at  ${\bf x}=0$ \cite{LandauElectro}:
\begin{equation*}
{\bf E}_{cl} = {\bf N}\frac{ne}{r^2}\frac{1-v^2}{(1-v^2\sin^2\theta)^{3/2}}, \  \
{\bf H}_{cl} = {\bf v}\times{\bf E}_{cl}.
\end{equation*}
The theory has enough structure to prove that it is the case, 
but as a check we have verified the result by a direct calculation to the order $\lambda^4$.
We also note that the action of the boost operator on the ground state
$|G\rangle$ additionally populates the $(\ell,m)$ modes in the $|G,\lambda\rangle$ state.
This allows for the addition of the non spherically-symmetric component to the
quantum Coulomb field, which is indispensable for the reproduction of the above classical result.

Second, it turns out that $|G\rangle$ is a 
squeezed vacuum state $|n\rangle$
(see Ref. \cite{Cohen} for the discussion of the squeezed states in the quantum optics context): 
\begin{equation}
|G\rangle = \exp(\hat D)|n\rangle, \
 \hat D = \frac{1}{2}\sum_{\ell=1}^\infty\sum_{m=-\ell}^\ell u_{\ell m}\left(\hat b^{\dag2}_{\ell m} - \hat b^2_{\ell m}\right).
\label{GS}
\end{equation}
This is easily seen from the following identity:
$
\exp(-\hat D)\hat B_{\ell m}\exp(\hat D) = \hat b_{\ell m}
$. Moreover,
the overlap between the ground state and the charged vacuum state  is
\begin{equation}
\langle n|G\rangle = \prod_{\ell=1}^\infty 
\left[\cosh\left(\frac{{\rm \ath}\left(\beta_\ell\right)}{2}\right)\right]^{-\ell-\frac{1}{2}}\approx0.997.
\nonumber
\end{equation}
The two spherically-symmetric states are therefore quite similar.
Interestingly, their overlap  is
independent of the fine structure constant $\alpha=e^2$. Thus, it is of purely 
geometric nature. Furthermore, we can study the population of the $(\ell,m)$ 
modes, 
\begin{equation}
\langle G|\hat b_{\ell m}^\dag\hat b_{\ell m}|G\rangle = 
\sh^2\left(\frac{\ath(\beta_\ell)}{2}\right)=\frac{\gamma_\ell}{4\sqrt{\ell(\ell+1)}}-\frac{1}{2}.
\label{bins}
\end{equation}
It  equals about $3\times10^{-3}$ for $\ell=1$, decays motonically with $\ell$, and for $\ell\gg1$ 
it approaches $1/64\ell^4$: only low $(\ell,m)$ modes are noticeably
populated. 

Third, while the expectation value of the field  operators (\ref{Efluc}) and
(\ref{Hfluc}) reproduces the classical result, their fluctuations in the ground
state have absolutely no classical counterpart.
This is so because the long-range fields are now described by a quantum field theory, and not just by the classical functions. 
Indeed, 
$$
\langle \hat {\bf E}(0,r,\theta,\varphi) \hat {\bf E}(0,r,\theta',\varphi') \rangle =
\frac{n^2e^2}{r^4}+\frac{1}{r^4} \sum_{\ell=1}^\infty
\frac{2\ell+1}{G_\ell^2}\,\eta_\ell\,
P_\ell\bigl(\cos\theta\cos\theta'+\cos(\varphi-\varphi')\sin\theta\sin\theta'\bigr),
$$
where $\eta_\ell=1$ when the average is calculated in the $n$-charged vacuum
state $|n\rangle$ and 
\begin{equation*}
\eta_\ell=\frac{G_\ell^2\sqrt{\ell(\ell+1)}}{2}>1
\end{equation*}
when it is 
calculated in the ground state $|G\rangle$. ($P_\ell$ stands for the Legendre polynomial.) This
shows explicitly that the charged states $|n\rangle$ and $|G\rangle$ (i) differ
by the pattern of the fluctuations of the electric field associated with them; (ii) are spherically symmetric. 

\section{Discussion of the results}
\label{sec_dis}
To start, this work proposes  that there can be a {\it fluctuating long-range} electromagnetic 
field around every charge. This field originates
from the quantum fluctuations of the Coulomb field. We propose that it should  
be considered in addition to the fluctuating vacuum field studied in the standard QED framework. 
Since
such a field has not been observed yet, the confirmation of its presence will
fundamentally update our understanding of charged particles.
It is thus critical to find out the experimentally accessible consequences of this {\it conjecture}.

General experience with quantum systems interacting  with the 
fluctuating electromagnetic fields is
that (at the very least) their energy levels are subjected to shifts \cite{MilonniBook1993}. 
We propose to
employ the spectroscopic measurements of the energy levels of the bound systems  such as hydrogen, 
muonic hydrogen, muonium, etc. in the hunt for  the 
experimental evidence of the quantum nature of the Coulomb field. 

To quantify the expected level shifts, one can adopt the perturbative formalism analogous 
to the one employed for the calculation of
the Lamb shift \cite{BethePhysRev,Bethe,MilonniBook1993,inprep}. 
In this approach the
quantum state of the Coulomb field of a nucleus is modeled by our ground state
$|G\rangle$ (\ref{GS}). The field operators used for writing down the coupling
between the orbiting charge and the fluctuating Coulomb field are given by 
Eqs. (\ref{Efluc}) and (\ref{Hfluc}). The excited states of the QCF and their energies (both required
in the perturbative expansion)
are provided by the eigenstates and  eigenvalues of  
our electromagnetic energy operator \eqref{Diag}. Since the eigenvalues depend on $\varepsilon_c$,
we propose to treat  it as the (only) free parameter and 
argue that its value can be experimentally determined from the spectroscopic measurements. 
Naturally, the larger $\varepsilon_c$ the smaller the level shift due to the QCF should be. 

The magnitude of the level shifts resulting from the presence of the QCF is 
bounded by the discrepancies between the experimental measurements  and the QED calculations.
For example, the frequency of the 1S-2S transition 
in hydrogen is experimentally known with the fantastic accuracy of about $40$ Hz corresponding
to the relative accuracy of about $1$ part in $10^{14}$ \cite{SpectroscopyAll,HaenschRMP2006}. Even more 
amazingly, this result is expected to be significantly 
improved in the foreseeable future \cite{HaenschRMP2006}. 
The QED calculations match it with 
the relatively thick error bar of several tens of kHz \cite{EidesBook2007}. 
The most of this theoretical uncertainty comes from the imprecise knowledge about the ``distribution of charge'' 
in a proton. Thus, the QCF corrections to this transition 
are bounded from above by a few tens of kHz.

The above-outlined possibility to consider additional level shifts in hydrogen-like atoms 
is especially appealing in the light of the current discrepancies between
the standard QED predictions  and 
the spectacular recent measurements of the Lamb shift in muonic hydrogen
\cite{PohlNature2010}. It is also quite natural  given the fact that 
precision spectroscopy  has been serving  for a long time as the Rosetta Stone for 
deciphering the laws of quantum physics \cite{HaenschRMP2006}. 
The formalism proposed in this manuscript should lay the ground for the studies
of the spectroscopic consequences of the quantum Coulomb field. The
work along these lines is already ongoing \cite{inprep}.

\begin{center}
\bf ACKNOWLEDGMENTS
\end{center}

This work is supported by U.S. Department of Energy through the LANL/LDRD Program  (BD)
and by the project MA4851/1-1 of the Deutsche Forschungsgemeinschaft (PM). 
We thank  Dr. Malcolm Boshier for drawing our attention to Ref.
\cite{PohlNature2010} and for useful discussions. We thank  Prof. Andrzej
Staruszkiewicz and Dr. Andrzej Herdegen for 
insightful comments about the theory of the quantum Coulomb field.

\appendix

\renewcommand{\thesection}{}
\section{}
\renewcommand{\theequation}{A\arabic{equation}}
\noindent We use spherical harmonics defined as 
\begin{equation}
Y_{\ell m}(\theta,\varphi) = (-)^{(m+|m|)/2} i^\ell  
\sqrt{\frac{2\ell+1}{4\pi}\frac{(\ell-|m|)!}{(\ell+|m|)!}} 
P^{|m|}_\ell(\cos\theta)\exp(im\varphi),
\label{Ylm}
\end{equation}
following  the convention  introduced  by Staruszkiewicz in Ref. \cite{Acta1995}.
For the convenience of the reader we list the first three $\ell$ sectors:
\begin{align*}
&Y_{00} =\frac{1}{\sqrt{4\pi}}, \ \ Y_{10} =i\sqrt{\frac{3}{4\pi}}\cos\theta, \\
&Y_{1,\pm1} =\mp i\sqrt{\frac{3}{8\pi}}\sin \theta\, e^{\pm i\vp}, \ \
Y_{20}  =\sqrt{\frac{5}{16\pi}}(1-3\cos^2 \theta), \\  
&Y_{2,\pm1}  =\pm \sqrt{\frac{15}{32\pi}}\sin(2 \theta) e^{\pm i\vp}, \ \
Y_{2,\pm 2} =-\sqrt{\frac{15}{32\pi}}\sin^2\theta e^{\pm 2i\vp}. 
\end{align*}
In particular, this implies that 
\begin{equation}
Y^*_{\ell m}(\theta,\varphi) = (-)^{m+\ell}Y_{\ell,-m}(\theta,\varphi), \ \
\hat L_\pm Y_{\ell m} = \sqrt{\ell(\ell+1)-m(m\pm1)}Y_{\ell,m\pm1},
\label{prop}
\end{equation}
where 
$$
\hat L_\pm = 
e^{\pm i\varphi}\left(\pm\frac{\partial}{\partial\theta} + i{\rm ctg}\theta \frac{\partial}{\partial\varphi}\right).
$$
In our manuscript, we use the redefined spherical harmonics $\tilde Y_{\ell m}$. 
Considering $m>0$, one has  
\begin{equation}
\tilde Y^*_{\ell m}(\theta,\varphi) = (-)^{m+\ell}\tilde Y_{\ell m}(\theta,\varphi), \ \ 
\tilde Y^*_{\ell,-m}(\theta,\varphi) = -(-)^{m+\ell}\tilde Y_{\ell,-m}(\theta,\varphi).
\label{conj_tildeY}
\end{equation}
Noting that 
$$
\sum_{m=-\ell}^\ell \tilde Y_{\ell m}(\theta,\varphi) \tilde Y^*_{\ell m}(\theta',\varphi') = 
\sum_{m=-\ell}^\ell Y_{\ell m}(\theta,\varphi) Y^*_{\ell m}(\theta',\varphi'),
$$
we obtain  
\begin{equation}
\sum_{m=-\ell}^\ell \tilde Y_{\ell m}(\theta,\varphi) \tilde Y^*_{\ell m}(\theta',\varphi') = 
\frac{2\ell+1}{4\pi} P_\ell(\cos\theta\cos\theta' +
\cos(\varphi-\varphi')\sin\theta\sin\theta'),
\label{ident}
\end{equation}
where $P_\ell$ is the Legendre polynomial.
 Eqs. (\ref{prop}), (\ref{conj_tildeY}) and (\ref{ident}) are useful in the derivation of 
both the electromagnetic energy operator and the expression for the two point correlation function.

\noindent The boost operator $\hat M^{03}$, that we use in the paper, was calculated
in Ref. \cite{Wrobel} (see also Ref. \cite{Rostworowski}). We list it below for reader's convenience:
\begin{equation}
\hat M^{03} = \frac{1}{\sqrt{6}\pi e}\hat Q(\hat c_{10}+\hat c_{10}^\dag) +   \frac{i}{4\pi
e^2}\sum_{\ell=2}^\infty\sum_{m=-\ell}^\ell
\sqrt{\frac{\ell^2-1}{4\ell^2-1}}\sqrt{\ell^2-m^2}(\hat c_{\ell m}^\dag \hat
c_{\ell-1, m} - {\rm h.c.}),
\label{AAA}
\end{equation}
which in our notation corresponds to 
\begin{equation}
\hat M^{03} =i\sqrt{\frac{2}{3\pi}}\hat Q(\hat b_{10}-\hat b^\dag_{10}) +  i
\sum_{\ell=2}^\infty\sum_{m=-\ell}^\ell
\sqrt{\frac{\ell^2-1}{4\ell^2-1}}\sqrt{\ell^2-m^2}(\hat b_{\ell m}^\dag \hat
b_{\ell-1, m} - {\rm h.c.}).
\label{AM03}
\end{equation}
We mention in passing that the derivation of the boost operators (\ref{AAA}) and
(\ref{AM03}) assumes that spherical harmonics are given by Eq. (\ref{Ylm}).


\begin{thebibliography}{32}
\bibitem[{foo({\natexlab{a}})}]{footnote1}
\bibinfo{note}{We note that the action of $\hat S_0$ on the vacuum state
  $|0\rangle$ is not defined in the theory. The only operators involving $\hat
  S_0$, whose action on $|0\rangle$ is defined, are the periodic functions of
  $\hat S_0$. This is in line with the well-known problems with the definition
  of the angle operator in Quantum Mechanics [see e.g. R. Peierls, {\it
  Surprises in theoretical physics} (Princeton University Press, 1979)].}
\bibitem[{foo({\natexlab{b}})}]{footnote2}
\bibinfo{note}{We note that there are also other techniques besides normal
  ordering for making the energy operator finite (see e.g. Ref.
  \cite{Parker_book} for the advanced tools of the modern theory of quantum
  fields in curved spacetime). They are expected to shift the zero point
  contribution given in Eq. (\ref{shift}), but do not affect the gap between
  eigenstates of the energy operator.}
\bibitem[{foo({\natexlab{c}})}]{footnote3}
\bibinfo{note}{Since an electron is structureless, the QCF theory should in
  principle describe its electromagnetic field on the arbitrary distance from
  its center. On the other hand, a proton is clearly characterized by some
  charge distribution absent in the QCF theory. As a result, the QCF theory
  should describe reliably the fluctuating electromagnetic field away from the
  proton charge core.}
\end{thebibliography}
\end{document}